\documentclass[12pt,aps,superscriptaddress,prl]{revtex4-1}  
\usepackage{graphicx}  
\usepackage{dcolumn}   
\usepackage{bm}        
\usepackage{amssymb}   
\usepackage{amsmath}   
\usepackage{amsthm}    
\usepackage{xcolor}    
\usepackage{tikz}      
\usepackage{ifthen}    
\usepackage{multirow}  
\usepackage{tabu}      
\usepackage{subfigure}
\usepackage{bbold}
\usepackage{xspace}
\usepackage[unicode=true,
linktocpage,
linkbordercolor={0.5 0.5 1},
citebordercolor={0.5 1 0.5},
linkcolor=blue]{hyperref}
\newcommand{\sro}{Sr$_2$RuO$_4$\xspace}
\newcommand{\Tc}{$T_\mathrm{c}$\xspace}
\newcommand{\Tcs}{$T_\mathrm{c}$s\xspace}

\begin{document}

\title{Strong Increase in Ultrasound Attenuation Below T$_\mathrm{c}$ in \sro: Possible Evidence for Domains}

\date{\today}
\author{Sayak Ghosh}
\affiliation{Laboratory of Atomic and Solid State Physics, Cornell University, Ithaca, NY 14853, USA}
\author{Thomas G. Kiely}
\affiliation{Laboratory of Atomic and Solid State Physics, Cornell University, Ithaca, NY 14853, USA}
\author{Arkady Shekhter}
\affiliation{National High Magnetic Field Laboratory, Florida State University, Tallahassee, FL 32310, USA}
\author{F. Jerzembeck}
\affiliation{Max Planck Institute for Chemical Physics of Solids, Dresden, Germany}
\author{N. Kikugawa}
\affiliation{National Institute for Materials Science, Tsukuba, Ibaraki 305-0003, Japan}
\author{Dmitry A. Sokolov}
\affiliation{Max Planck Institute for Chemical Physics of Solids, Dresden, Germany}
\author{A. P. Mackenzie}
\affiliation{Max Planck Institute for Chemical Physics of Solids, Dresden, Germany}
\affiliation{SUPA, School of Physics and Astronomy, University of St Andrews, North Haugh, St Andrews KY16 9SS, UK}
\author{B.~J.~Ramshaw}
\email{bradramshaw@cornell.edu}
\affiliation{Laboratory of Atomic and Solid State Physics, Cornell University, Ithaca, NY 14853, USA}	
\maketitle

\newpage

\textbf{Recent experiments suggest that the superconducting order parameter of \sro has two components. A two-component order parameter has multiple degrees of freedom in the superconducting state that can result in low-energy collective modes or the formation of domain walls---a possibility that would explain a number of experimental observations including the smallness of the time reversal symmetry breaking signal at \Tc and telegraph noise in critical current experiments. We perform ultrasound attenuation measurements across the superconducting transition of \sro using resonant ultrasound spectroscopy (RUS). We find that the attenuation for compressional sound increases by a factor of seven immediately below \Tc, in sharp contrast with what is found in both conventional ($s$-wave) and high-\Tc ($d$-wave) superconductors. We find our observations to be most consistent with the presence of domain walls between different configurations of the superconducting state. The fact that we observe an increase in sound attenuation for compressional strains, and not for shear strains, suggests an inhomogeneous superconducting state formed of two distinct, accidentally-degenerate superconducting order parameters that are not related to each other by symmetry. Whatever the mechanism, a factor of seven increase in sound attenuation is a singular characteristic with which any potential theory of the superconductivity in \sro must be reconciled.}

\section*{Introduction}

One firm, if perhaps counter-intuitive, prediction of Bardeen, Cooper, and Schrieffer (BCS) theory is the contrasting behavior of the nuclear spin relaxation rate, $1/T_1$, and the ultrasonic attenuation, $\alpha$ \cite{BCS1957}. Upon cooling from the normal state to the superconducting (SC) state, one might expect both $1/T_1$ and $\alpha$ to decrease as both processes involve the scattering of normal quasiparticles. In the SC state, however, Cooper pairing produces correlations between quasiparticles of opposite spin and momentum. These correlations produce ``coherence factors'' that add constructively for nuclear relaxation and produce a peak---the Hebel-Slichter peak---in $1/T_1$ immediately below \Tc \cite{SlichterPRL1957}. In contrast, the coherence factors add destructively for sound attenuation and there is an immediate drop in $\alpha$ below \Tc \cite{MorsePRL1959}. These experiments provided some of the strongest early evidence for the validity of BCS theory \cite{BCS1957}, and the drop in sound attenuation below \Tc was subsequently confirmed in many elemental superconductors \cite{LevyPRL1963,LeibowitzPRL1964,ClaibornePRL1965,FossheimPRL1967}.

It came as a surprise, then, when peaks in the sound attenuation were discovered below \Tc in two heavy-fermion superconductors: UPt$_3$ and UBe$_{13}$ \cite{BatloggPRL1985,GoldingPRL1985,MullerSolStComm1986}. Specifically, peaks were observed in the longitudinal sound attenuation---when the sound propagation vector $\mathbf{q}$ is parallel to the sound polarization $\mathbf{u}$: ($\mathbf{q}\parallel\mathbf{u}$). Transverse sound attenuation ($\mathbf{q}\perp\mathbf{u}$), on the other hand, showed no peak below \Tc but instead decreased with power law dependencies on $T$ that were ultimately understood in terms of the presence of nodes in the SC gap \cite{MorenoPRB1996}. Various theoretical proposals were put forward to understand the peaks in the longitudinal sound attenuation, including collective modes, domain-wall friction, and coherence-factors \cite{MiyakePRL1986,MonienSol1987,JoyntPRL1986,CoffeyPRB1986}, but the particular mechanisms for UPt$_3$ and UBe$_{13}$ were never pinned down (see \citet{SigristRMP1991} for a review). What is clear, however, is that a peak in sound attenuation below \Tc is not a prediction of BCS theory and surely indicates unconventional superconductivity.

\begin{figure*}
	\centering
	\includegraphics[width=0.99\linewidth]{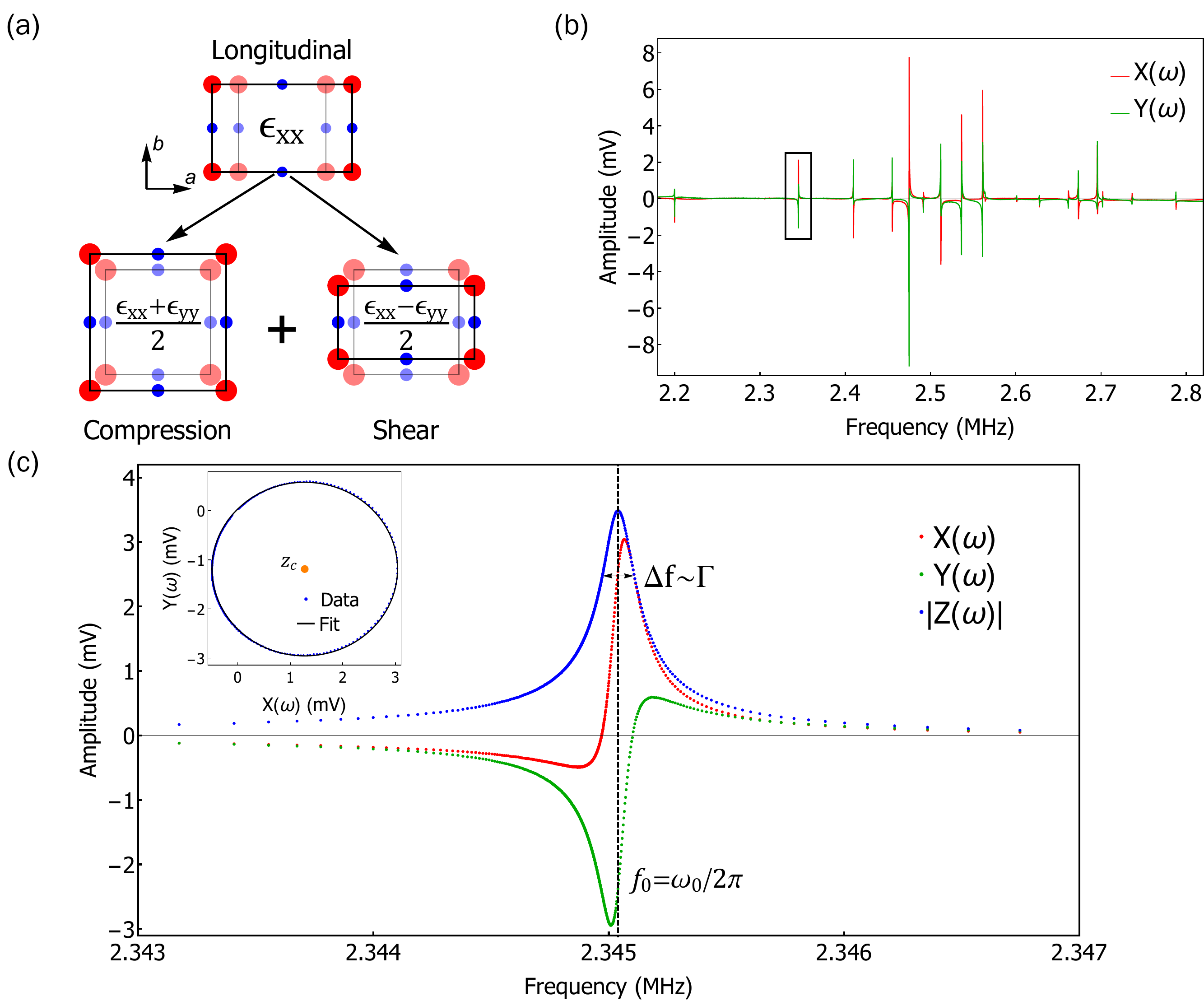}
	\caption{\label{fig:spectra} Measuring ultrasonic attenuation with resonant ultrasound spectroscopy. (a) The \sro unit cell under a deformation corresponding to the longitudinal strain $\epsilon_{xx}$, associated with the elastic constant $c_{11}$. This mode is a superposition of pure compression $\epsilon_{xx}+\epsilon_{yy}$ and pure shear $\epsilon_{xx}-\epsilon_{yy}$, associated with the elastic constants $(c_{11}+c_{12})/2$ and $(c_{11}-c_{12})/2$, respectively. (b) Resonant ultrasound spectrum of \sro between 2.2-2.8 MHz. $X(\omega)$ and $Y(\omega)$ are the real and the imaginary parts of the response. The boxed resonance is shown in detail in (c). (c) Zoom-in on the resonance near 2.34 MHz. The center of the resonance and the linewidth are indicated. Inset shows the same resonance plotted in complex plane and fit to a circle---$z_c$ denotes the center of the circle.} 
	
\end{figure*}

The superconductivity of \sro has many unconventional aspects, including time reversal symmetry (TRS) breaking \cite{luke1998time,xia2006high,KidwingiraScience2006}, the presence of nodal quasiparticles \cite{LupienPRL2001,HassingerPRX2017,SharmaPNAS2020}, and two components in its SC order parameter \cite{GhoshNatPhys2020,BenhabibNatPhys2020}. These observations have led to various recent theoretical proposals for the SC state in \sro \cite{AlinePRB2019,romer2019knight,RoisingPRR2019,ScaffidiArxiv2020,SuhPRR2020,KivelsonNPJ2020,WillaPRB2021}, requiring further experimental inputs to differentiate between them. Not only should the coherence factors differ for \sro compared to the $s-$wave BCS case, but there is the possibility of low-energy collective modes \cite{chung2012charge} and domain-wall motion \cite{Yuan2021}, all of which could be observable in the ultrasonic attenuation.


Prior ultrasonic attenuation measurements on \sro reported a power-law temperature dependence of the transverse sound attenuation, interpreted as evidence for nodes in the gap \cite{LupienPRL2001}, but found no other unconventional behaviour. This may be in part due to the specific ultrasound technique employed: pulse-echo ultrasound. While pulse-echo can measure a pure shear-strain response in the transverse configuration, the in-plane longitudinal configuration is a combination of both compression strain and shear strain in a tetragonal crystal like \sro \cite{Brugger1965}. Specifically, the L100 mode measures the elastic constant $c_{11}$, which is a mixture of pure compression, $(c_{11}+c_{12})/2$, and pure shear, $(c_{11}-c_{12})/2$ (see \autoref{fig:spectra}(a)). Shear and compression strains couple to physical processes in fundamentally different ways and thus effects that couple exclusively to compressional sound may have been missed in previous measurements.



\section*{Experiment}

We have measured the ultrasonic attenuation of \sro across \Tc using an ultrasound technique distinct from pulse-echo ultrasound---resonant ultrasound spectroscopy (RUS). RUS allows us to obtain the attenuation in all the independent symmetry channels in a single experiment (i.e. for all 5 irreducible representations of strain in \sro). See \citet{GhoshNatPhys2020} for details of our custom built low-temperature RUS apparatus. The high-quality \sro crystal used in this experiment was grown by the floating zone method---more details about the sample growth can be found in \citet{BobowskiCondMat2019}. A single crystal was precision-cut along the [110], [1$\bar{1}$0] and [001] directions and polished to the dimensions 1.50~mm $\times$ 1.60~mm $\times$ 1.44~mm,  with 1.44~mm along the tetragonal $c$ axis. The sample quality was characterized by heat capacity and AC susceptibility measurements, as reported in \citet{GhoshNatPhys2020}. The SC \Tc measured by these techniques---approximately 1.43 K---agrees well with the \Tc seen in our RUS experiment, indicating that the sample underwent uniform cooling during the experiment.

RUS measures the mechanical resonances of a three-dimensional solid. The frequencies of these resonances depend on the elastic moduli, density, and geometry of the sample, while the widths of these resonances are determined by the ultrasonic attenuation \cite{RamshawPNAS,GhoshSciAdv2020}. Because each resonance mode is a superposition of multiple kinds of strain, the attenuation in all strain channels can be extracted by measuring a sufficient number of resonances---typically 2 or 3 times the number of unique strains (of which there are 5 for \sro). 

A typical RUS spectrum from our \sro sample is shown in \autoref{fig:spectra}(b) (see methods for details of the measurement.) Each resonance can be modeled as the response $Z(\omega)$ of a damped harmonic oscillator driven at frequency $\omega$ (see \autoref{fig:spectra}(c)),
\begin{equation}
	Z(\omega)=X(\omega)+iY(\omega)= Ae^{i\phi}/((\omega-\omega_0)+i\Gamma)
\end{equation}
where $X$ and $Y$ are the real and imaginary parts of the response, and $A$, $\Gamma$, and $\phi$ are the amplitude, linewidth, and phase, respectively. The real and imaginary parts of the response form a circle in the complex plane. The response is measured at a set of frequencies that space the data points evenly around this circle: this is the most efficient way to precisely determine the resonant frequency $\omega_0$ and the linewidth $\Gamma$ in a finite time (see \citet{ShekhterNat2013} for details of the fitting procedure). We plot the temperature dependence of the linewidth of all our experimentally measured resonances through \Tc in the SI. For comparison, the attenuation $\alpha$ measured in conventional pulse-echo ultrasound is related to the resonance linewidth via $\alpha = \Gamma/v$, where $v$ is the sound velocity.


\section*{Results}

When the sound wavelength, $\lambda = \frac{2\pi}{q}$, is much longer than the electronic mean free path $l$, i.e. when $ql\ll1$, the electron-phonon system is said to be in the `hydrodynamic' limit \cite{Khan1987} (this is different than the hydrodynamic limit of electron transport). Given that the best \sro has a mean free path that is at most of order a couple of microns, and that our experimental wavelengths are of the order of 1 mm, we are well within the hydrodynamic limit. In this regime, we can express the linewidth $\Gamma$ of a resonance $\omega_0$ as,
\begin{equation}
	\frac{\Gamma}{\omega_0^2}=\frac{1}{2}\sum_{j}\alpha_j\frac{\eta_j}{c_j},
	\label{eqn:visc}
\end{equation} 
where $\eta_j$ and $c_j$ are the independent components of the viscosity and elastic moduli tensors, respectively (see SI for details). The $\alpha$ coefficients define the composition of a resonance, such that $\alpha_j=\partial(\ln \omega_0^2)/\partial(\ln c_j)$ and $\sum_{j}\alpha_j=1 $\cite{RamshawPNAS}.



\begin{figure}
	\centering
	\includegraphics[width=0.65\linewidth]{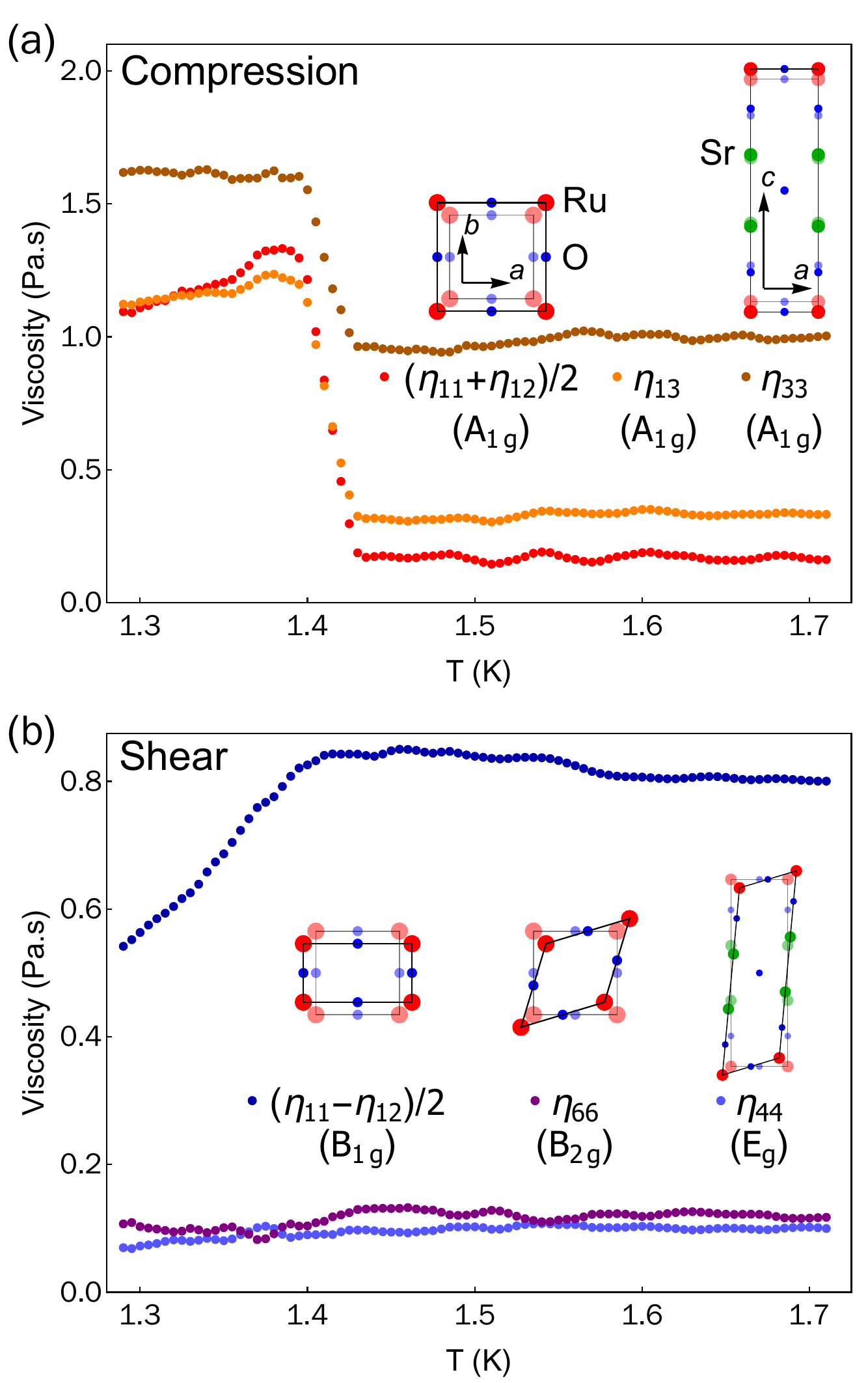}
	\caption{\label{fig:visc} Symmetry-resolved sound attenuation in \sro. (a) Compressional and (b) shear viscosities through \Tc. The irreducible strain corresponding to each viscosity is shown---$\eta_{13}$ arises due to coupling between the two $A_{1g}$ strains. An increase in the compressional viscosities is seen immediately below \Tc, while no such feature is seen in the shear viscosities.}
\end{figure}

We measured the linewidths of 17 resonances and resolved them into the independent components of the viscosity tensor. The tetragonal symmetry of \sro dictates that there are only six independent components, arising from the five irreducible representations (irreps) of strain in $D_{4h}$ plus one component arising from coupling between the two distinct compression strains \cite{GhoshNatPhys2020}. The six symmetry-resolved components of viscosity in \sro are plotted in \autoref{fig:visc}. 

The shear viscosity $(\eta_{11}-\eta_{12})/2$ decreases below \Tc in a manner similar to what is observed in conventional superconductors \cite{MorsePRL1959,LevyPRL1963}. We find that $(\eta_{11}-\eta_{12})/2$ is much larger than the other two shear viscosities, which is consistent with previous pulse-echo ultrasound experiments \cite{LupienPRL2001,LupienThesis}. On converting attenuation to viscosity, we find relatively good agreement (within a factor of 2) between the resonant ultrasound and pulse-echo measurements of $(\eta_{11}-\eta_{12})/2$. This is particularly non-trivial given that sound attenuation scales as frequency squared in the hydrodynamic regime and the pulse-echo ultrasound measurements were performed at frequencies roughly two orders of magnitude higher than those used in the RUS measurements. The much larger magnitude of $(\eta_{11}-\eta_{12})/2$, in comparison to $\eta_{66}$, may be due to the fact that the $\epsilon_{xx}-\epsilon_{yy}$ strain is associated with pushing the $\gamma$ Fermi surface pocket toward the van Hove singularity \cite{barber2019role}. The small values of $\eta_{44}$ and $\eta_{66}$ are comparable to the experimental noise and any changes at \Tc are too small to resolve at these low frequencies.

In contrast with the rather conventional shear viscosities, the three compressional viscosities each exhibit a strong increase below \Tc. For in-plane compression---the strain that should couple strongest to the rather two-dimensional superconductivity of \sro---the increase is by more than a factor of seven. After peaking just below \Tc, the attenuation slowly decreases as the temperature is lowered. The large increase below \Tc was not observed in previous longitudinal sound attenuation measurements made by pulse-echo ultrasound \cite{LupienPRL2001,LupienThesis}. Longitudinal sound is a mixture of pure shear and pure compression, as shown in \autoref{fig:spectra}(a). At the frequencies where pulse-echo ultrasound is measured---of order 100 MHz---the shear viscosity $(\eta_{11}-\eta_{12})/2$ is at least one order of magnitude larger than the compression viscosity \cite{LupienThesis} and thus completely dominates the sound attenuation. The relative time-scales between the dynamics of the attenuation mechanism and the measurement frequency may also play a role---we will return to this idea later on in the discussion. 

\section*{Analysis}


\begin{figure*}
	\centering
	\includegraphics[width=0.99\linewidth]{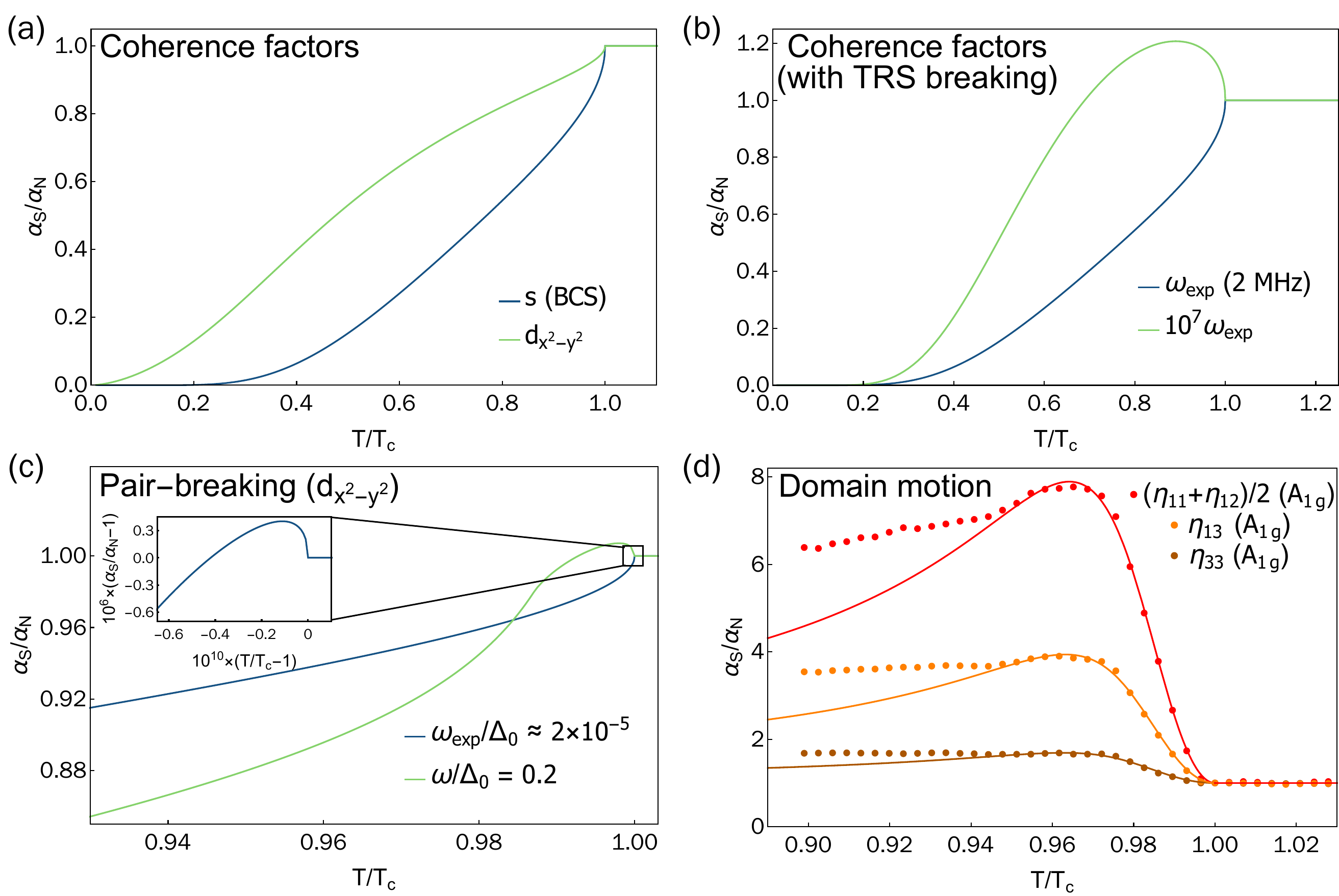}
	\caption{\label{fig:fits} Comparison of different mechanisms for sound attenuation in the superconducting state. (a) Attenuation in the superconducting state $\alpha_S$ (normalized by the normal state attenuation $\alpha_N$) for an isotropic $s$-wave gap and a $d_{x^2-y^2}$ gap, calculated within the BCS framework. (b) $\alpha_S/\alpha_N$ for a time reversal symmetry breaking gap below \Tc. A peak is seen at high enough frequencies ($\sim$THz) but not at our experimental frequencies ($\sim$MHz). (c) Attenuation peak at different frequencies due to pair-breaking effects in a $d_{x^2-y^2}$ gap. The inset shows the plot at our experimental frequency in detail---a tiny peak is seen about 0.01 nK below \Tc. (d) Normalized attenuation in  the $A_{1g}$ channels of \sro through \Tc, fit to the attenuation expected from domain wall motion below \Tc. The fit works well only close to \Tc, probably because it does not include other temperature-dependent effects (see text for details).}
\end{figure*}


We now analyze possible mechanisms that could give rise to such an increase in sound attenuation below \Tc. First, we calculate sound attenuation within a BCS-like framework, accounting for the differences in coherence factors that occur for various unconventional SC order parameters. We find that a peak can indeed arise under certain circumstances but not under our experimental conditions. Second, we consider phonon-induced Cooper pair breaking in the SC state that does lead to a sound attenuation peak just below \Tc, but which is inaccessibly narrow in our experiment. Finally, we show that a simple model of sound attenuation due to the formation of SC domains best matches the experimental data. 


First we examine the possibility of increased sound attenuation due to coherent scattering in the SC state. Sound attenuation and nuclear spin relaxation in an $s$-wave superconductor are proportional to the coherence factors $F_{\pm} = (1\pm \Delta_0^2/E_k E_{k'})$, where $\Delta_0$ is the uniform $s$-wave gap and $E_k$ is the Bogoliubov quasiparticle dispersion \cite{BCS1957}. Scattering off of a nucleus flips the spin of the quasiparticle and the resultant coherence factor is $F_+$, where the $+$ sign produces the Hebel-Slichter peak below \Tc. Scattering off of a phonon, on the other hand, does not flip quasiparticle spin and the resultant coherence factor is $F_-$, producing a sharp drop in sound attenuation below \Tc. In general, the coherence factors depend on the structure of the superconducting gap, motivating the idea that an unconventional superconducting OP might produce a peak in the sound attenuation. Calculating within the BCS framework, we find that attenuation for a $d_{x^2-y^2}$ gap decays slowly compared to the isotropic $s$-wave gap, but does not exhibit a peak (\autoref{fig:fits}(a), see SI for details of the calculation). This slow decrease can be attributed to the presence of nodes in the $d_{x^2-y^2}$ gap \cite{Maki1994,Vekhter1999}. For a TRS breaking gap such as $p_x+ip_y$ or $d_{xz}+id_{yz}$, a Hebel-Slichter-like peak appears below \Tc if sufficiently large-angle scattering is allowed (\autoref{fig:fits}(b)). Scattering at these large wavevectors---essentially scattering across the Fermi surface---would require ultrasound with nanometer wavelengths. This regime is only accessible at THz frequencies, whereas our experiment operates in the MHz range. Hence we rule out coherent scattering as the mechanism of increased compressional sound attenuation below \Tc.

Next we consider how phonon-induced Cooper pair breaking may give rise to a sound attenuation peak, similar to what has been observed in superfluid $^3$He-$B$ below \Tc \cite{AdenwallaPRL1989}. Pair-breaking in BCS superconductors requires a minimum energy of 2$\Delta_0$, where $\Delta_0$ is the gap magnitude. This energy scale is generally much higher than typical ultrasound energies. For example, the maximum gap magnitude in \sro is 2$\Delta\sim$ 0.65 meV \cite{FirmoPRB2013}, which would require a frequency of approximately 1 THz to break the Cooper pairs. However, the pair-breaking energy is lowered for a gap with nodes, such as $d_{x^2-y^2}$. In particular, since the gap goes to zero at \Tc, it may be small enough near \Tc such that pair-breaking is possible at a few MHz. Our calculations for a $d_{x^2-y^2}$ gap, however, show that $\sim$10 GHz frequencies are required to produce an experimentally discernible peak (\autoref{fig:fits}(c)). At our experimental frequencies, the peak is only visible within 0.01 nK of \Tc. For a fully gapped superconductor, like the TRS breaking state $p_x + i p_y$, the peak will be even smaller. This clearly rules out pair-breaking as the origin of the increased sound attenuation.

Finally, we consider the formation of SC domains. When different configurations of a SC order parameter are degenerate, such as $p_x + i p_y$ and $p_x - i p_y$, domains of each configuration will form separated by domain walls. These domain walls can oscillate about their equilibrium positions when sound propagates through the sample \cite{JoyntPRL1986}. \citet{SigristRMP1991} derive an expression for the sound attenuation coefficient, $\alpha$, from domain wall motion of the form
\begin{equation}
\alpha\left(\omega,T\right) \propto \frac{\omega^2}{\omega^2+\omega_{DW}^2}\rho_s^2,
\label{eq:atten1}
\end{equation}
where $\rho_s$ is the superfluid density (proportional to the square of the superconducting gap), $\omega$ is the angular frequency of the sound wave, and $\omega_{DW}$ is the lowest vibrational frequency of the domain wall. While the full functional form of $\rho_s$ and $\omega_{DW}$ are unknown, near \Tc they can be written within a Ginzburg-Landau (GL) formalism as $\rho_s\propto|T-T_c|$ and $\omega_{DW}\propto|T-T_c|^{3/2}$ which gives the explicit temperature dependence of the above equation as 
\begin{equation}
\alpha\left(\omega,T\right) \propto \frac{\omega^2}{\omega^2+\omega_1^2\left|T/T_{\rm c}-1\right|^3}\left|T/T_{\rm c}-1\right|^2,
\label{eq:atten}
\end{equation}
where $\omega_1$ is the domain wall frequency as $T\rightarrow 0$. We fit all three attenuation channels to \autoref{eq:atten} and extract $\omega_1 = 500\pm25$ MHz (\autoref{fig:fits}(d)). As the temperature approaches \Tc from below the domain wall frequency decreases to zero, producing a peak in the attenuation when the experimental frequency is approximately equal to the domain wall frequency. This may partially explain why no peak was observed in previous pulse-echo ultrasound measurements at $\sim$100 MHz \cite{LupienPRL2001} (as mentioned above, there is also the issue that the attenuation for longitudinal sound along the [100] direction is almost entirely dominated by $(\eta_{11}-\eta_{12})/2$).) The fit of \autoref{eq:atten} deviates substantially from the data for $T/T_{\rm c}\lesssim 0.95$: this is to be expected, as the GL theory is only valid near \Tc \cite{SigristRMP1991}. Nevertheless, \autoref{eq:atten} captures the correct shape of the rapid increase in attenuation below \Tc in all three compression channels, using the same value of $\omega_1$ for all three fits. The extracted frequency scale of $\omega_1 \approx 500$ MHz is also reasonable: studies of sound attenuation in nickel at MHz frequencies show similar magnitudes of increase in the magnetically ordered state when domains are present \cite{Leonard1962}. We note that the results of Josephson interferometry measurements have previously been interpreted as evidence for SC domains in \sro \cite{KidwingiraScience2006}.

\section*{Discussion}

The factor of seven increase we find in the in-plane compressional viscosity is without precedent in a superconductor. For comparison, longitudinal attenuation increases by 50\% below \Tc in UPt$_3$ \cite{MullerSolStComm1986}, and by a bit more than a factor of two in UBe$_{13}$ \cite{GoldingPRL1985}. There is also a qualitative difference between the increase in \sro and the increase seen in the heavy fermion superconductors: the attenuation peaks sharply below \Tc in both UPt$_3$ and UBe$_{13}$, with a peak width of approximately 10\% of \Tc. The compressional attenuation in \sro, by contrast, decreases by only about 10\% over the same relative temperature range. This suggests that something highly unconventional occurs in the SC state of \sro, leading to a large increase in sound attenuation that is not confined to temperatures near \Tc. The mechanism we find most consistent with the data is domain wall motion.

Assuming that we have established the likely origin of the increase in sound attenuation, we consider its implications for the superconductivity of \sro. The formation of domains requires a two-component order parameter (OP), either symmetry-enforced or accidental, reaffirming the conclusions of recent ultrasound studies of the elastic moduli and the sound velocity \cite{GhoshNatPhys2020,BenhabibNatPhys2020}. We can learn more about which particular OPs are consistent with our experiment by considering which symmetry channels show the increase in attenuation. Domains attenuate ultrasound when the application of strain raises or lowers the condensation energy of one domain in comparison to a neighboring domain. A simple example is the ``nematic'' superconducting state proposed by \citet{BenhabibNatPhys2020}, which is a $d-$wave OP of the $E_g$ representation, transforming as $\left\{d_{xz},d_{yz}\right\}$. Under $(\epsilon_{xx}-\epsilon_{yy})$ strain, domains of the $d_{xz}$ configuration will be favored over the $d_{yz}$ configuration (depending on the sign of the strain). This will cause some domains to grow and others to shrink, attenuating sound through the mechanism proposed by \citet{SigristRMP1991}. We find no increase in $\left(\eta_{11}-\eta_{12}\right)/2$ below \Tc, suggesting that a $\left\{d_{xz},d_{yz}\right\}$ OP cannot explain the increase in compressional sound attenuation.

More generally, the lack of increase in attenuation in any of the shear channels implies that that the SC state of \sro does not break rotational symmetry. Domains that are related to each other by time reversal symmetry can also be ruled out: there is no strain that can lift the degeneracy between, for example, a $p_x + i p_y$ domain and a $p_x - i p_y$ domain. The observed increase in sound attenuation  under compressional strain is therefore quite unusual: as \citet{SigristRMP1991} point out, compressional strains can never lift the degeneracy between domains that are related by \textit{any} symmetry, since compressional strains do not break the point group symmetry of the lattice. Instead, attenuation in the compressional channel requires domains that couple differently to compressional strain, which in turn requires domains that are accidentally degenerate. Examples that are consistent with both NMR \cite{PustogowNat2019} and ultrasound \cite{GhoshNatPhys2020,BenhabibNatPhys2020} include $\{d_{x^2-y^2},g_{xy(x^2-y^2)}\}$ \cite{KivelsonNPJ2020,WillaPRB2021,ClepkensArxiv2021} and $\{s,d_{xy}\}$ \cite{RomerArXiv2021}. Then, for example, domains of $d_{x^2-y^2}$ will couple differently to compressional strain than domains of $g_{xy(x^2-y^2)}$, leading to the growth of one domain type and an increase in compressional sound attenuation below \Tc. Shear strain, meanwhile, does not change the condensation energy of any single-component order parameter (e.g. $s$, $d_{xy}$, $d_{x^2-y^2}$, or $g_{xy(x^2-y^2)}$) to first order in strain, which means that the lack of increase in shear attenuation below \Tc is also consistent with an accidentally-degenerate OP. This is also consistent with the lack of a cusp in \Tc under applied shear strain \cite{HicksSc2014,watson2018micron}.

Recent theoretical work \cite{Yuan2021} has shown that domain walls between $d_{x^2-y^2}$ and $g_{xy(x^2-y^2)}$ OPs may provide a route to explain the observation of half-quantum vortices in \sro \textit{without} a spin-triplet order parameter \cite{Jang2011}---a result that is otherwise inconsistent with the singlet pairing suggested by NMR \cite{PustogowNat2019}. \citet{WillaPRB2021}, followed by \citet{Yuan2021},  have shown that domains between such states stabilize a TRS-breaking $d_{x^2-y^2} \pm i g_{xy(x^2-y^2)}$ state near the domain wall. This would naturally explain why probes of TRS breaking, such as the Kerr effect and $\mu$SR \cite{xia2006high,GrinenkoNatPhys2021}, see such a small effect at \Tc in \sro. 

One significant challenge for the two-component order parameter scenario is that, whether accidentally degenerate or not, a two component order parameter should generically produce two superconducting \Tcs. The lack of a heat capacity signature from an expected second transition under uniaxial strain \cite{LiPNAS2021} can only be explained if the second, TRS-breaking transition is particularly weak---a result that might be consistent with the TRS-breaking state appearing only along domain walls. Finally, it is worth noting that there are other mechanisms of ultrasonic attenuation that we have not explored here, including collective modes and gapless excitations such as edge currents that might appear along domain walls even if the domains are related by symmetry. Future ultrasound experiments under applied static strain and magnetic fields are warranted as certain types of domain walls can couple to these fields, thereby affecting the sound attenuation through \Tc.

\section*{Acknowledgments}
B. J. R. and S. G. acknowledge support for building the experiment, collecting and analyzing the data, and writing the manuscript from the Office of Basic Energy Sciences of the United States Department of Energy under award no. DE-SC0020143.  B.J.R. and S.G. acknowledge support from the Cornell Center for Materials Research with funding from the Materials Research Science and Engineering Centers program of the National Science Foundation (cooperative agreement no. DMR-1719875). T. G. K. acknowledge support from the National Science Foundation under grant no. PHY-2110250.  N. K. acknowledges support from Japan Society for the Promotion of Science (JSPS) KAKENHI (Nos. JP17H06136, JP18K04715, and 21H01033)) and Japan Science and Technology Agency Mirai Program (JPMJMI18A3) in Japan.

\newpage

\section*{Supplemental Material}
	
	\subsection*{Viscosity from RUS Measurements}
	
\begin{figure*}
		\centering
		\includegraphics[width=.99\linewidth]{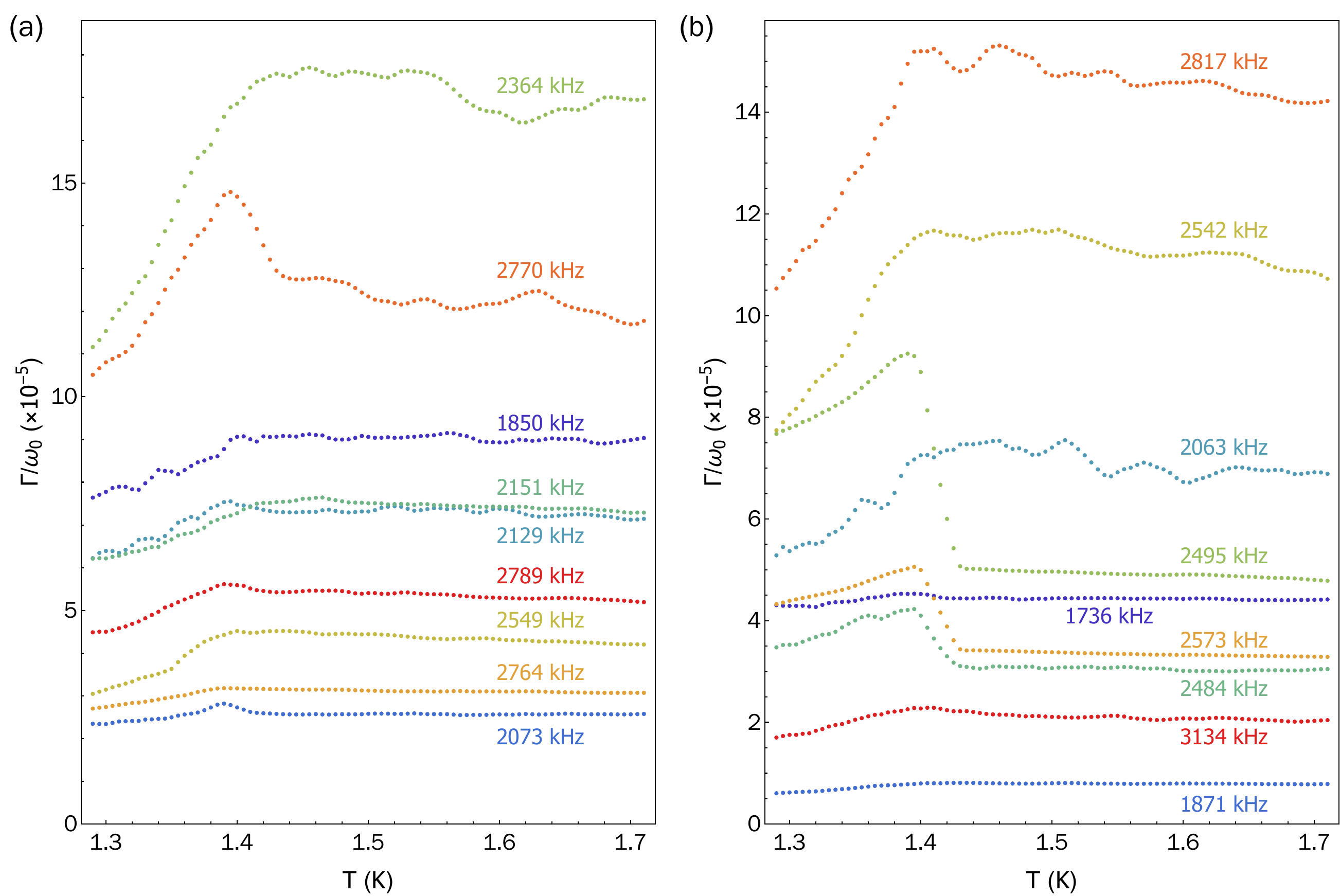}
		\caption{\textbf{RUS attenuation data.} (a) Temperature evolution of normalized resonance linewidth of 18 resonances of \sro through \Tc, with panels (a) and (b) each showing 9 resonances. These 18 resonances were used to calculate the six independent components of the viscosity tensor through \Tc.}
		\label{fig:gammavT}
	\end{figure*}

	RUS measures the mechanical resonances of a solid, which are determined by the elastic constants and the density of the material along with its dimensions. Each resonance $\omega_0$ is a superposition of the various irreducible strains, and therefore is a function of the independent elastic moduli $c_j$. In a tetragonal system like \sro, there are six such moduli ($j=1,2,...,6$) such that
	\begin{equation}
	\omega_0^2=\mathfrak{F}(c_j,\rho,l_k),
	\label{eqn:freqdef}
	\end{equation}
	where $\rho$ is the density and $l_k$ are the dimensions of the sample. Sound attenuation in the solid leads to these frequencies having a finite linewidth $\Gamma$, giving them a characteristic Lorentzian shape (as discussed in the main). Within linear response, sound attenuation is related to the strain rate through the viscosity tensor \cite{MorenoPRB1996}, which has the same symmetries as the elastic moduli tensor. To relate the experimentally measured linewidths to the irreducible viscosities, we replace $\omega_0\rightarrow\omega_0+i\Gamma$ and $c_j\rightarrow c_j+i\omega_0\eta_j$ in \autoref{eqn:freqdef},
	\begin{equation}
	\begin{aligned}
	&(\omega_0+i\Gamma)^2=\mathfrak{F}(c_j+i\omega_0\eta_j,\rho,l_k)\\
	\implies &\omega_0^2 + 2i\omega_0\Gamma \approx \mathfrak{F}(c_j,\rho,l_k)+\sum_{j}\frac{\partial \mathfrak{F}}{\partial c_j}\cdot i\omega_0\eta_j\\
	\implies&\Gamma =\frac{1}{2}\sum_{j}\frac{\partial \mathfrak{F}}{\partial c_j}\cdot \eta_j =\frac{1}{2}\omega_0^2\sum_{j}\alpha_j \frac{\eta_j}{c_j},
	\end{aligned}
	\label{eqn:gammadef}
	\end{equation}
	where $\alpha_j=\partial(\ln \omega_0^2)/\partial(\ln c_j)$ and $\sum_{j}\alpha_j=1$ \cite{RamshawPNAS}. The fit procedure outlined in \citet{RamshawPNAS} gives us the $\alpha_j$ coefficients for all our experimentally measured resonances. Knowing these coefficients, we can calculate the six independent viscosities as a function of temperature by measuring the temperature evolution of sufficiently many resonance linewidths (typically 2-3 times the number of independent viscosities). Note that \autoref{eqn:gammadef} is true in the weak attenuation limit ($\Gamma\ll\omega_0$), which is easily satisfied in our experiments ($\Gamma/\omega_0\sim10^{-4}$ for all our measured resonances, see \autoref{fig:gammavT}).

	
	
	\subsection*{Quasiparticle Scattering in the Superconducting State}
	
	The conventional method for describing ultrasound attenuation assumes that sound waves attenuate by scattering quasiparticles. In particular, we assume that the attenuation rate is proportional to the scattering rate induced by the sound wave. The effect of the superconducting transition on scattering was introduced by BCS~\cite{BCS1957} and is described pedagogically by Tinkham~\cite{Tinkham} and Schrieffer~\cite{Schrieffer}. We assume that the induced scattering can be described by an interaction Hamiltonian
	\begin{equation}
	\mathcal{H}_{\rm int}=\sum_{k,k^\prime,\sigma}M_{k,k^\prime}c^\dagger_{k,\sigma}c_{k^\prime,\sigma}
	\end{equation}
	where $M_{k,k^\prime}$ is symmetric under time reversal (TR). This is crucial: as an $s$-wave superconductor consists of time-reversed fermionic pairs, the symmetry of the interaction under TR has drastic effects on the scattering properties. In particular, an interaction Hamiltonian that is even under TR will result in destructive interference between Bogoliubov quasiparticles, and vice versa.
	
	The scattering rate can be computed by Fermi's golden rule. Following Tinkham~\cite{Tinkham}, we take $M_{k,k^\prime}=Me^{i\theta_{k,k^\prime}}$ to have a constant modulus. This is a drastic approximation that allows for arbitrarily wide-angle scattering. While it has little effect on the $s$-wave calculation, we will have to remedy this approximation for the $p$-wave gap function. We perform the calculations in 2D for each of the Fermi surfaces of ${\rm Sr}_2{\rm RuO}_4$. The band structure adds a Jacobian factor to the integrand, $J_i(\phi)$, where $\phi$ is the azimuthal angle in momentum space and $i\in\{\alpha,\beta,\gamma\}$ for the $\alpha$, $\beta$ and $\gamma$ bands, respectively. The general integral is of the form
	\begin{multline}
	\Gamma_s(\omega,T)=|M|^2\int_0^\infty dE~\int\frac{d\phi_1d\phi_2}{(2\pi)^2}J_i(\phi_1)J_i(\phi_2)n_s(E,\phi_1)n_s(E+\hbar\omega,\phi_2)\times\\\big(f(E)-f(E+\hbar\omega)\big)\big(F_{\gamma^\dagger\gamma}(E,\phi_1,\phi_2)+F_{\gamma\gamma}(E,\phi_1,\phi_2)\big)
	\label{eq:gammaS}
	\end{multline}
	where $n_s(E,\phi)$ is the angle-resolved density of states, obeying $\int \frac{d\phi}{2\pi}n_s(E,\phi)=N_s(E)$, $f(E)$ is the Fermi factor, and $F(E,\phi_1,\phi_2)$ is the coherence factor. The coherence factors arise due to interference between quasiparticles. They are derived by rewriting the interaction Hamiltonian in terms of Bogoliubov quasiparticles:
	\begin{equation}
	\mathcal{H}_{\rm int}=M\sum_{k,k^\prime}F_{\gamma^\dagger\gamma}(k,k^\prime)\gamma^\dagger_{k,\sigma}\gamma_{k^\prime,\sigma}+M\sum_{k,k^\prime}F_{\gamma\gamma}(k,k^\prime)\gamma_{k,\sigma}\gamma_{k^\prime,\sigma}+h.c.
	\end{equation}
	Linearizing the functions $F(k,k^\prime)$ about the Fermi surface and evaluating the energy delta function gives us the functions $F(E,\phi,\phi^\prime)$.
	
	The coherence factors depend on the structure of the gap function in spin and momentum space. We will not treat the general case here.  
	For the case of a $p_x+ip_y$ gap function, the coherence factors are both of the form
	\begin{equation}
	F_\pm(E,\phi,\phi^\prime)=1\pm\frac{\Delta_0^2}{E(E+\hbar\omega)}\cos(\phi-\phi^\prime).
	\end{equation}
	In Figure~3 we plot the results for the $(p_x+ip_y)\hat{z}$ gap function, for which $F_{\gamma^\dagger\gamma}=F_{\gamma\gamma}=F_+$.
	As we are considering a 2D scattering problem, the gap function has a constant modulus in momentum space, so the angle-resolved density of states is equal to the density of states:
	\begin{equation}
	n_s(E,\phi)=N_s(E)=\frac{1}{\sqrt{E^2-\Delta_0^2}}.
	\end{equation}
	Collecting terms, we find that the scattering rate is given by
	\begin{multline}
	\Gamma_{x+iy}(\omega,T)=2|M|^2\int_0^\infty dE~\int\frac{d\phi_1d\phi_2}{(2\pi)^2}J_i(\phi_1)J_i(\phi_2)\frac{1}{\sqrt{(E^2-\Delta_0^2)((E+\hbar\omega)^2-\Delta_0^2)}}\times\\\big(f(E)-f(E+\hbar\omega)\big)\bigg(1-\frac{\Delta_0^2}{E(E+\hbar\omega)}\cos(\phi_1-\phi_2)\bigg).
	\end{multline}
	This is essentially equivalent to the s-wave scattering problem except for the factor of $\cos(\phi_1-\phi_2)$. Importantly, one can immediately see that the integral over $J_i(\phi_1)J_i(\phi_2)\cos(\phi_1-\phi_2)$ will vanish due to the $\mathbb{C}_4$ symmetry of the band structure. This means that the coherence factors are effectively equal to 1, as if the Bogoliubov quasiparticles do not interfere with one another. It is this lack of interference that leads to the peak shown in panel (b) of Fig.~3. 
	
	Note that the vanishing (average) interference between quasiparticles depends on integrating the relative angle, $\phi_1-\phi_2$, over the full period of the cosine function. This involves wide-angle scattering events, where the particles being scattered sit on opposite sides of the Fermi surface. Such scattering events would require that the sound wave have a momentum ${\bf q}={\bf k_1}-{\bf k_2}$ that can be as large as $2k_F$. Experimentally, however, the sound waves have a frequency $q/2k_F\sim 10^{-7}$. We therefore propose a phenomenological method for confining the relative angle $|\phi_1-\phi_2|\lesssim\eta$ by including the Gaussian factor $G(\phi_1-\phi_2)$ in the integrand, where
	\begin{equation}
	G(x)=\frac{1}{\eta\sqrt{2\pi}}e^{-(x/\eta)^2/2}.
	\end{equation}
	As the relative angle on a circular Fermi surface is given by $\phi_1-\phi_2=\arcsin(q/2k_F)$, we convert the parameter $\eta$ to an equivalent sound frequency using $\omega_\eta=2v_sk_F\sin(\eta)$. For large $\omega_\eta$, we recover the non-interfering results. For $\omega_\eta$ on the order of the experimental frequency, however, $\cos(\phi_1-\phi_2)\approx 1$ and we recover the standard s-wave calculation with a pronounced dip across $T_c$. This corresponds to strong destructive interference between quasiparticles. These curves are compared in panel (b) of Fig.~3.
	
\begin{figure}
		\centering
		\includegraphics[width=.65\linewidth]{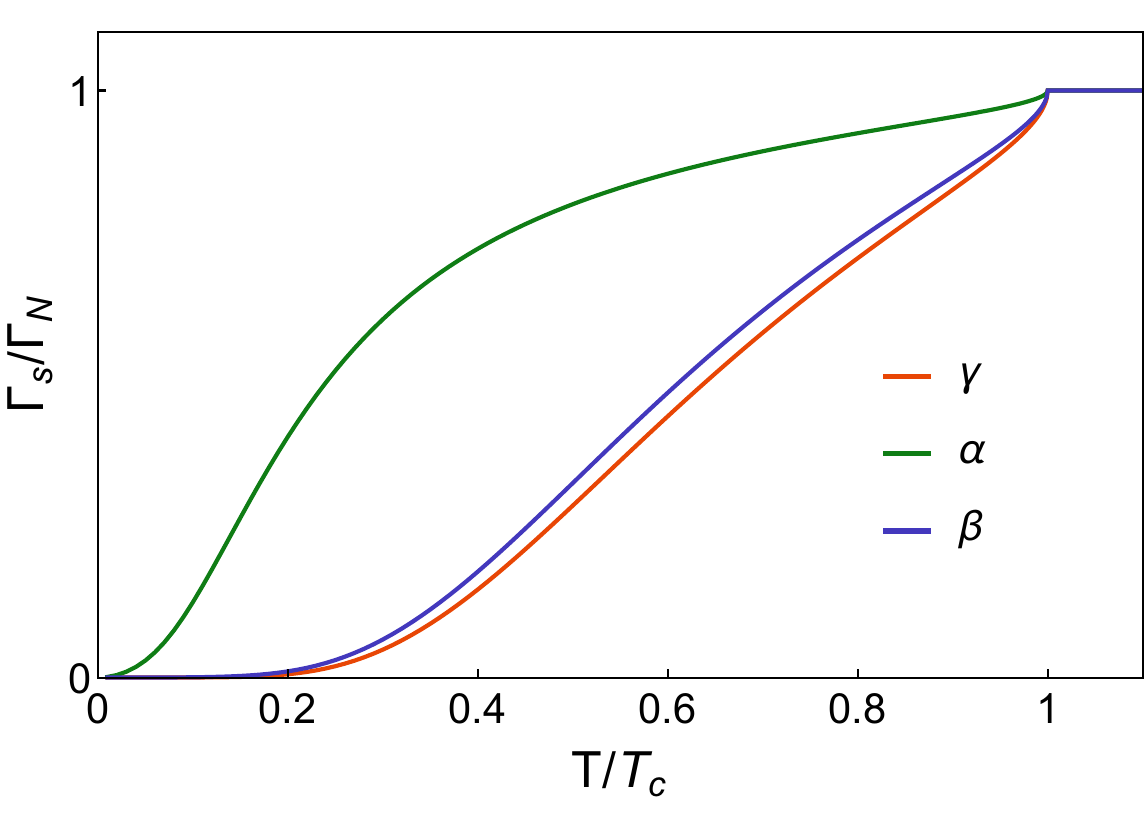}
		\caption{{\bf Scattering rate for ${\rm Sr}_2{\rm RuO}_4$ bands.} Comparison of the scattering rate with a $d_{x^2-y^2}$ gap function for each of the three ${\rm Sr}_2{\rm RuO}_4$ Fermi surfaces. The electron-like $\gamma$ and $\beta$ bands show nearly identical behavior while the hole-like $\alpha$ band differs significantly at low temperatures. We find that these discrepancies do not change the qualitative behavior across $T_c$. Note that the $\beta$ curve is reproduced in panel (a) of Fig.~3 in the main text.}
		\label{fig:dWvFig}
	\end{figure}

	The coherence factors for a $d_{x^2-y^2}$ gap function are of the form
	\begin{equation}
	F_\pm(E,\phi,\phi^\prime)=1\pm\frac{\Delta_0^2}{E(E+\hbar\omega)}|\cos(2\phi)||\cos(2\phi^\prime)|
	\end{equation}
	and the angle-resolved density of states is
	$n_s(E,\phi)=|E|/\sqrt{E^2-\Delta_0^2\cos^2(2\phi)}$. Inserting these directly into \autoref{eq:gammaS}, one obtains
	\begin{multline}
	\Gamma_{x^2-y^2}(\omega,T)=2|M|^2\int_0^\infty dE~\int\frac{d\phi_1d\phi_2}{(2\pi)^2}J_i(\phi_1)J_i(\phi_2)\big(f(E)-f(E+\hbar\omega)\big)\times\\\frac{1}{\sqrt{(E^2-\Delta_0^2\cos^2(2\phi_1))((E+\hbar\omega)^2-\Delta_0^2\cos^2(2\phi_2))}}\bigg(1-\frac{\Delta_0^2}{E(E+\hbar\omega)}|\cos(2\phi_1)||\cos(2\phi_2)|\bigg).
	\label{eq:dWaveEq}
	\end{multline}
	We do not include the phenomenological Gaussian factor for the results in panel (a) of Fig.~3 because there is no erroneous peak, so we do not expect any qualitative change in the results. In principle, however, the structure of the $d$-wave gap in momentum space means that a quantitative calculation of the scattering rate should remove wide-angle scattering events. All theory curves in panels (a), (b) and (c) of Fig.~3 were computed for the $\beta$ band. \autoref{fig:dWvFig} shows the normalized scattering rate for the $\alpha$, $\beta$, and $\gamma$ bands.
	The $\alpha$ band is hole-like, and its scattering rate differs qualitatively from that of the electron-like bands. We note that this effect appears most pronounced at low temperatures and that it does not change the behavior across $T_c$. The $s$-wave curve, also plotted in panel (a) of Fig.~3, has coherence factors $1\pm\frac{\Delta_0^2}{E(E+\hbar\omega)}$ and a density of states $n_s(E)=(E^2-\Delta_0^2)^{-1/2}$. The $s$-wave singlet problem is covered extensively in Refs.~\cite{BCS1957,Tinkham,Schrieffer} so we do not review it here.
	
	\subsection*{Cooper Pair-breaking below \Tc}
	
	In \autoref{eq:gammaS}, we neglected a set of quadratic Bogoliubov terms that are referred to as pair-breaking terms. These terms are neglected because the domain of the integral formally vanishes when the driving frequency $\omega<2\Delta(T)$ where $\Delta(T)$ is the temperature-dependent s-wave gap. For gap functions with nodes, however, the frequency domain does not vanish explicitly. Crucially, the pair breaking coherence factors have the opposite sign of the dominant coherence factors discussed above. This means that an interaction that leads to destructive interference between quasiparticles (and thus a sharp drop in the scattering rate across $T_c$) will exhibit constructive interference in its pair-breaking terms. Thus, pair breaking terms will induce a peak where there is otherwise none.
	
	Pair breaking terms can be identified immediately by the energy delta function. For example, in going from the general statement of Fermi's golden rule to \autoref{eq:gammaS}, we evaluated the energy delta function $\delta(E-E^\prime-\hbar\omega)$ assuming that $E,E^\prime>0$. This is the natural choice in the $\omega\to 0$ limit. For $\omega>0$, however, there are also solutions to the delta function where $E<0$ ($E^\prime>0$). These contributions are what we will explicitly include now. Note that even with nodes in the gap function, the domain of this integral formally vanishes as $\omega\to 0$. 
	
	For the d-wave gap, normal contributions are given by \autoref{eq:dWaveEq}. Pair breaking contributions differ in the domain of the $E$ integral, the sign of the coherence factor, and in the product of Fermi factors:
	\begin{multline}
	\Gamma^{PB}_{x^2-y^2}(\omega,T)=2|M|^2\int_0^\omega dE~\int\frac{d\phi_1d\phi_2}{(2\pi)^2}J_i(\phi_1)J_i(\phi_2)\big(1-f(E)-f(E+\hbar\omega)\big)\times\\\frac{1}{\sqrt{(E^2-\Delta_0^2\cos^2(2\phi_1))((E+\hbar\omega)^2-\Delta_0^2\cos^2(2\phi_2))}}\bigg(1+\frac{\Delta_0^2}{E(E+\hbar\omega)}|\cos(2\phi_1)||\cos(2\phi_2)|\bigg).
	\end{multline}
	We sum these two terms, normalized by the scattering rate in the normal state, for two values of $\omega$ in panel (c) of Fig.~3. An interpretation of these results is covered in the main text.
	
	\begin{figure*}
		\centering
		\includegraphics[width=.7\linewidth]{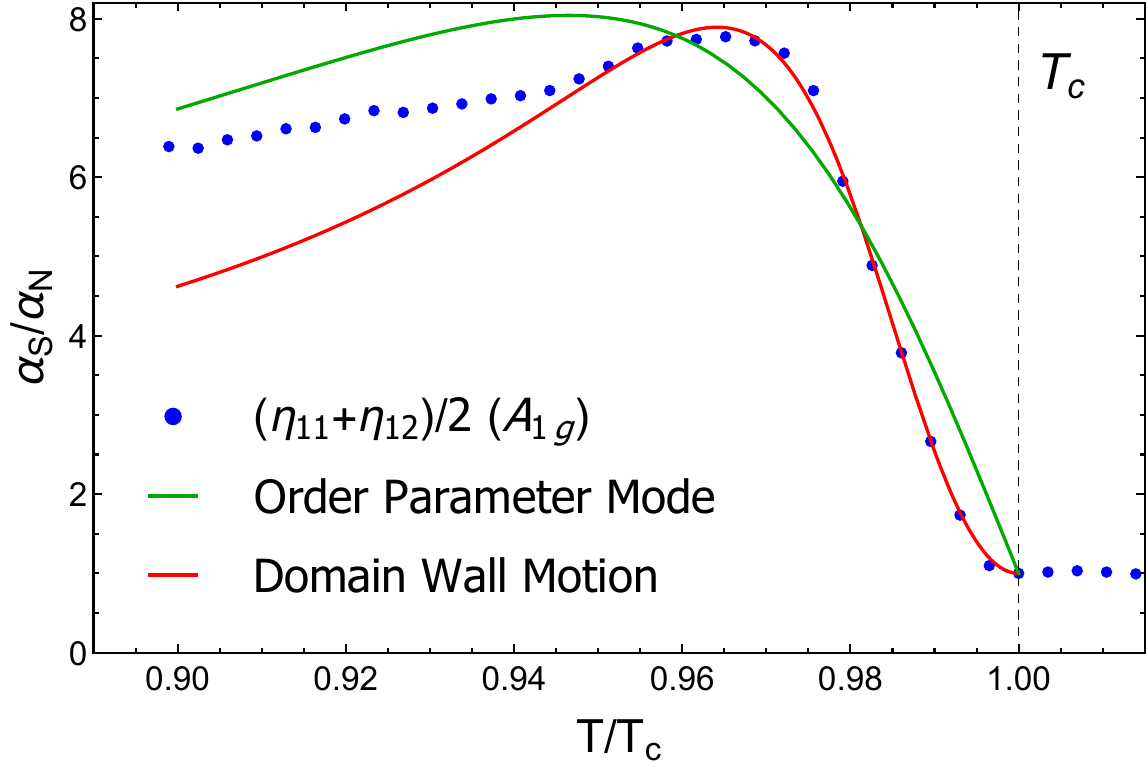}
		\caption{\textbf{Sound attenuation from order parameter modes.} Normalized $(\eta_{11}+\eta_{12})/2$ in \sro fit to two different models of increased sound attenuation below \Tc. The green curve is a fit to \autoref{eqn:att1}, which models sound attenuation due to OP modes. The red curve is a fit to \autoref{eqn:att2}, which models the sound attenuation arising from domain wall motion. Near \Tc, the red curve clearly fits the experimental data better than the green one.}
		\label{fig:compare}
	\end{figure*} 
	
	\subsection*{Attenuation Due to Order Parameter Relaxation}
	
	The formation of the SC order parameter below \Tc can lead to relaxational dynamics as the OP interacts with the strain. Within a Landau theory, the relaxation timescale diverges as $\left|T/T_{\rm c}-1\right|^{-1}$ close to \Tc. Unlike the resonant sound absorption arising from domains, OP relaxation can cause non-resonant absorption of ultrasound and lead to a broad peak in sound attenuation below \Tc \cite{SigristPTP2002}. We fit our measured $(\eta_{11}+\eta_{12})/2$ to the attenuation expression derived by \citet{SigristPTP2002},
	\begin{equation}
	\alpha(\omega,T)\propto \frac{\omega^2\tau}{1+\omega^2\tau^2} \sim \frac{\omega^2\tau_0/\left|T/T_{\rm c}-1\right|}{1+\omega^2\tau_0^2/\left|T/T_{\rm c}-1\right|^2}
	\label{eqn:att1}
	\end{equation}
	as shown in \autoref{fig:compare}. However we find that this expression does not capture the sharp increase in attenuation below \Tc, which the expression for attenuation from domain wall motion given in \citet{SigristRMP1991} does, 
	\begin{equation}
	\alpha\left(\omega,T\right) \propto \frac{\omega^2}{\omega^2+\omega_1^2\left|T/T_{\rm c}-1\right|^3}\left|T/T_{\rm c}-1\right|^2.
	\label{eqn:att2}
	\end{equation}
	In fact, the sharp peak-like behavior of attenuation right below \Tc, which is already present in the raw data (for example, 2495 kHz and 2573 kHz in \autoref{fig:gammavT}), points strongly to a resonant absorption mechanism compared to a non-resonant one.

\end{document}